\documentclass[USenglish,twocolumn]{article}

\usepackage[utf8]{inputenc}
\usepackage[big]{dgruyter}

\begin{document}

  \articletype{Research Article{\hfill}Open Access}

  \author*[1]{V. V. Bobylev}

\author[2]{A. T. Bajkova}

  \affil[1]{Pulkovo Observatory, E-mail: vbobylev@gao.spb.ru}

  \affil[2]{Pulkovo Observatory, E-mail: bajkova@gao.spb.ru}

  \title{\huge OB Stars and Cepheids From the Gaia TGAS Catalogue: Test of their Distances and Proper Motions}

  \runningtitle{OB Stars and Cepheids From the Gaia TGAS Catalogue}

  %\subtitle{...}

\begin{abstract}
{We consider young distant stars from the Gaia TGAS catalog. These are 250
classical Cepheids and 244 OB stars located at distances
up to 4~kpc from the Sun. These stars are used to determine the Galactic rotation parameters
using both trigonometric parallaxes and proper motions of the TGAS stars. In this case
the considered stars have relative parallax errors less
than 200\%. Following the well-known statistical approach, we assume
that the kinematic parameters found from the line-of-sight velocities
$V_r$ are less dependent on errors of distances than
the found from the velocity components $V_l$.
From values of the first derivative of the Galactic
rotation angular velocity $\Omega{'}_0$, found from the
analysis of velocities $V_r$ and $V_l$ separately, the scale factor of distances is determined.
We found that from the sample of Cepheids the scale of distances of the TGAS should be
reduced by 3\%, and from the sample of OB stars, on the contrary, the scale should be
increased by 9\%.}
\end{abstract}


\journalname{Open Astronomy}
\DOI{DOI}
  \startpage{1}
  \received{..}
  \revised{..}
  \accepted{..}

  \journalyear{2017}
  \journalvolume{4}
%  \journalissue{1}

\maketitle

\section{Introduction}

The Gaia TGAS (the Tycho-Gaia astrometric solution, Prusti et al.
2016; Lindegren et al. 2016) catalogue was produced from a
combination of data in the first year of Gaia observations with
Hipparcos/Tycho~(1997) stellar positions. The mean parallax
errors are $\sim0.3$~mas. This means that a solar neighborhood
with a radius of $\sim$300~pc can be covered with distance errors
of about 10\%. Therefore, when studying the structure and
kinematics of the Galaxy at great heliocentric distances (3~kpc or
more), the approach using the photometric or other distance scales
is currently topical. For most TGAS stars the mean proper motion
error is $\sim1$~mas yr$^{-1}$, but for quite a few Hipparcos
stars this error is an order of magnitude smaller, $\sim0.06$ mas
yr$^{-1}$ (Brown et al. 2016). Therefore, analyzing their space
velocities using highly accurate proper motions is of great
interest.

Comparison of distances to Cepheids and RR Lyra variables from
the TGAS catalog with the distances obtained by other
methods, showed very good agreement up to
distances 2~kpc (Casertano et al. 2017, Benedict et al.
2017; Clementini et al. 2017; Bobylev~\&~Bajkova, 2017). But
there are some reports about a zero-point offset of parallaxes of the
TGAS stars. In particular, Stassun~\&~Torres (2016) discovered the
offset equal to $-0.25\pm0.05$~mas with respect to 158 calibration
eclipsing binary stars. From the analysis of the nearest stars to the Sun
(<25~pc) Jao et al. (2017) found that, on average, the parallaxes of the Gaia
TGAS stars are less by $0.24\pm0.02$~mas than trigonometric parallaxes of these stars measured by ground-based telescopes.

Since the properties of the distance scale of the TGAS catalog are not yet studied completely,
the analysis of it with the use of independent approaches is actual.
The purpose of this paper is to determine the Galactic rotation parameters
using high-precision proper motions of the TGAS stars, a joint and separate solution of the main kinematic
equations, both in terms of proper motions and line-of-sight
velocities in order to investigate the distance scale of the TGAS catalog.

\section{Method}
From observations we know three components of the star velocity:
the line-of-sight velocity $V_r$ and the two projections of the
tangential velocity $V_l=4.74 r\mu_l\cos b$ and $V_b=4.74 r\mu_b,$
directed along the Galactic longitude $l$ and latitude $b$
respectively and expressed in km s$^{-1}$. Here, the coefficient
4.74 is the ratio of the number of kilometers in astronomical unit
to the number of seconds in a tropical year, and $r$ is a
heliocentric distance of the star in kpc. The components of a
proper motion of $\mu_l\cos b$ and $\mu_b$ are expressed in the
mas yr$^{-1}$.

To determine the parameters of the Galactic rotation curve, we use
the equations derived from Bottlinger's formulas in which the
angular velocity $\Omega$ was expanded in a series to terms of the
second order of smallness in $r/R_0:$
\begin{equation}
 \begin{array}{lll}
 V_r=-U_\odot\cos b\cos l-V_\odot\cos b\sin l\\
 -W_\odot\sin b+R_0(R-R_0)\sin l\cos b\Omega{'}_0\\
 +0.5R_0(R-R_0)^2\sin l\cos b\Omega^{''}_0,
 \label{EQ1}
 \end{array}
 \end{equation}
 \begin{equation}
 \begin{array}{lll}
 V_l= U_\odot\sin l-V_\odot\cos l-r\Omega_0\cos b\\
 +(R-R_0)(R_0\cos l-r\cos b)\Omega{'}_0\\
 +0.5(R-R_0)^2(R_0\cos l-r\cos b)\Omega^{''}_0,
 \label{EQ2}
 \end{array}
 \end{equation}
where $R$ is the distance from the star to the Galactic rotation
axis,
  \begin{equation}
 R^2=r^2\cos^2 b-2R_0 r\cos b\cos l+R^2_0.
 \end{equation}
$\Omega_0$ is the angular velocity of Galactic rotation at the
solar distance $R_0$, the parameters $\Omega{'}_0$ and
$\Omega^{''}$ are the corresponding derivatives of the angular
velocity, and $V_0=|R_0\Omega_0|$. The Oort constants $A$ and $B$
can be found from the following expressions:
\begin{equation}
  A=-0.5\Omega{'}_0 R_0,\quad B=-\Omega_0+A, \label{AB}
\end{equation}
written so that the following relations are hold: $A-B=\Omega_0$ and
$A+B=-(\Omega_0+\Omega{'}_0 R_0)$. In this paper, it is customary
the value of $R_0=8.0\pm0.2$~kpc, which Vall\'ee (2017) in its
recent survey found as the most probable.
%%%%%%%%%%%%%%%%%%%%%%%%%%%%%%%%%%%%%%%
%%%%%%%%%%%%%%%%%%%%%%%%%%%%%%%%%%%%%%%
There exists also the third Bottlinger's equation with the $V_b$
velocity on the left-hand side. The results of the analysis of
Cepheids and OB stars on the basis of a system of three
conditional equations with $V_r, V_l, V_b$ left-hand sides are
described in the papers of Bobylev (2017) and Bobylev~\&~Bajkova
(2017). However, for distant young stars with the latitudes close
to zero  ($b\approx0^\circ,\Rightarrow\sin b\approx0$) using the
equation with  $V_b$ left-hand side is ineffective. Therefore, in
this paper we either solve a system of two
equations~(\ref{EQ1})--(\ref{EQ2}), or separately each of them. It
is known (Zabolotskikh et al. 2002), that in such an approach the
velocity $W_\odot$ can not be reliably determined only from
equation~(\ref{EQ1}), so we fix its value as $W_\odot=-7$~km s
$^{-1}$.

The values $\Omega{'}_0$, obtained with separate solutions, are of great interest for controlling the distance scale. This method is based on the fact that the errors of the line-of-sight velocities are independent of
the distance errors, but the errors of the tangential components of the proper motion depend on the distance errors.
This is the basis for statistical methods for analyzing the distance scale.
For example, the distance scale factor $p$ can be found from
comparison of the values  $\Omega{'}_0$ obtained by different methods
(Zabolotskikh et al. 2002, Rastorguev et al. 2017), or from comparison of
spatial velocities $U,V,W$ or their variances
$\sigma_U, \sigma_V, \sigma_W$ (Sch\"onrich \& Aumer, 2017).

%%%%%%%%%%%%%%%%%%%%%%%%%%%%%%  FIGURE 1
\begin{figure}[t]
  \includegraphics[width=80mm]{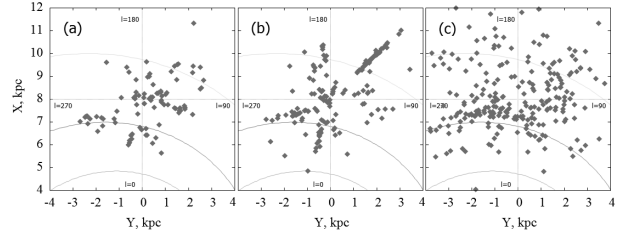}
 \caption{
The distribution on the Galactic plane $XY$ of spectral-double
OB stars (a), OB stars with distances determined from interstellar Calcium lines
(b), and Cepheids (c); the Sun has coordinates
$(X,Y)=(8,0)$~kpc; the four-armed spiral pattern with a pitch angle
$-13^\circ $ is plotted according to Bobylev \& Bajkova~(2014).
 }
 \label{f1}
 \end{figure}
%%%%%%%%%%%%%%%%%%%%%%%%%%%%%%%%%%%%%%%%

\section{Data}
In the present work we use the following two samples of OB stars:
1) the sample of spectroscopic binary OB stars (Bobylev \&
Bajkova~2015), 2) the sample of OB stars with distances determined
by the spectral lines of the interstellar Ca\,II (Bobylev \&
Bajkova~2011), and 3)the sample of classical Cepheids described in
Mel'nik et al.~(2015). The latter consists of 290 classical Cepheids with distances,
line-of-sight velocities and proper motions from the
Gaia TGAS catalog.

The distributions on the Galactic plane $XY$ of stars of three
considered samples are given in figure~\ref{f1}. As it can be seen
from this figure, all the stars are distant on no more than 5~kpc
from the Sun. OB stars are very young, therefore they well trace a
spiral pattern. It is also seen that among the older Cepheids,
there are a sufficient number of stars well tracing the
Carina-Sagittarius arm ($R\approx7$~kpc).

%%%%%%%%%%%%%%%%%%%%%%%%%%%%%%%%%%%%%%%%%%%%%%%%%%%%%%%%%%%%%% t1
 \begin{table}[t]
 \caption[]
  {\small
Parameters of Galactic rotation, found from Cepheids with the TGAS proper motions and trigonometric parallaxes
  }
  \begin{center}  \label{t:01}
  \small
  \protect
  \begin{tabular}{|l|r|r|r|r|r|}
   %\hline
   Parameter                  &      $V_r,V_l$  &          $V_r$  &      $V_l$  \\\hline
   $U_\odot$            & $ 7.5\pm1.0$    & $ 7.3\pm1.6$    & $ 8.6\pm1.0$  \\
   $V_\odot$            & $12.5\pm1.1$    & $13.6\pm1.4$    & $ 6.3\pm1.4$  \\

 $\Omega_0$         & $29.13\pm0.50$  &            ---  & $27.44\pm0.83$  \\
 $\Omega{'}_0$   & $-4.04\pm0.15$  & $-4.03\pm0.20$  & $-4.16\pm0.19$  \\
 $\Omega_0^{''}$  & $ 0.97\pm0.11$  & $ 0.81\pm0.16$  & $ 0.74\pm0.08$  \\
   $\sigma_0$           & $       14.31$  &      $  14.34$  &       $11.71$  \\
               $N_\star$       & $        220 $  &      $    216$  &        $240 $ \\
           $A$        & $ 16.17\pm0.60$ & $ 16.11\pm0.78$ & $ 16.66\pm0.78$ \\
           $B$        & $-12.96\pm0.78$ &             --- & $-10.78\pm1.14$ \\
 %\hline
 \end{tabular}\end{center}
 {\small Note: $U_\odot, V_\odot, \sigma_0$ are in km s$^{-1}$, $\Omega_0, A, B$ are in km s$^{-1}$ kpc$^{-1}$,
 $\Omega{'}_0$ is km s$^{-1}$ kpc$^{-2}$,  $\Omega_0^{''}$ is in km s$^{-1}$ kpc$^{-3}$.}
 \end{table}
%%%%%%%%%%%%%%%%%%%%%%%%%%%%%%%%%%%%%%%%%%%%%%%%%%%%%%%%%%%%%%%%% t1

\section{Results and Discussion}\label{Results}
\subsection{Cepheids}\label{Ceph}
In table~\ref{t:01} the parameters of Galactic rotation obtained from Cepheids with
the TGAS proper motions and parallaxes are given. We used the Cepheids with relative errors
of trigonometric parallaxes $\sigma_\pi/\pi<200\%$. In the table the error of unit weight $\sigma_0, $ obtained by solving conditional equations~(\ref{EQ1})--(\ref{EQ2}) using a well-known least squares method is given. This error is close to the  dispersion of residual velocities of the analyzed sample stars averaged over all directions. The Oort constants $A$ and $B$ calculated using (\ref{AB}) are also given in the table. According to this table, we find the coefficient of the distance scale $p=(-4.03)/(-4.16)=0.97\pm0.07$. The error of $p$ was calculated
as follows
 $$
 \sigma^2_p=(\sigma_{\Omega'_{0r}}/\Omega'_{0t})^2+
          (\Omega'_{0r}\sigma_{\Omega'_{0t}}/\Omega'^2_{0t})^2,
          $$
where $\Omega'_{0t}$ is $\Omega'_{0(l,b)}$. Such an estimate from
Cepheids was obtained, apparently, for the first time.

It is possible to estimate the distance at which the accuracy of
the tangential $V_t$ velocity is equal to the accuracy of the
line-of-sight velocity $V_r.$ Using the upper bound of the
accuracy estimate of the tangential component, from the equality
 $\sigma_{V_t}=4.74r\sqrt{\sigma^2_{\mu_{\alpha}\cos\delta}+\sigma^2_{\mu_\delta}}$
 we find the value of the critical distance, near
which the tangential velocities are more accurate than the
line-of-sight ones. Based on the fact that $\sigma_{V_r}=3$~km
s$^{-1}$ and
$\sigma_{\mu_{\alpha}\cos\delta}=\sigma_{\mu_\delta}=1$~mas
yr$^{-1}$ for TGAS stars we obtain $r=0.45$~kpc. Therefore, it is
most correct to apply our method as follows: for distances greater
than the critical value, it is necessary to consider the kinematic
parameters obtained from the line-of-sight velocities of the stars
to be more accurate, and at distances less than the critical one,
it is necessary to consider the kinematic parameters obtained from
the proper motions of the stars to be more accurate.

It is important to note that in our sample average distance
${\overline r}=2$~kpc. Only 9 Cepheids are located closer than
0.45~kpc from the Sun. Their exclusion from the sample does not
affect an estimate of the coefficient $p.$

Using the distances to Cepheids, calculated on the basis of
the period-luminosity relation and the proper motions of
the TGAS catalog, the following kinematic
parameters were obtained from joint solution of equations~(\ref{EQ1})--(\ref{EQ2}):
\begin{equation}
 \begin{array}{ccc}
 (U,V)_\odot=(7.9,11.0)\pm(0.8,1.0)~{\hbox {\rm km s$^{-1}$}},\\
     \Omega_0=29.87\pm0.45~{\hbox {\rm km s$^{-1}$ kpc$^{-1}$}},\\
  \Omega{'}_0=-4.13\pm0.13~{\hbox {\rm km s$^{-1}$ kpc$^{-2}$}},\\
 \Omega_0^{''}=0.50\pm0.12~{\hbox {\rm km s$^{-1}$ kpc$^{-3}$.}}
 \label{Ceph1}
 \end{array}
 \end{equation}
Here we used 250 Cepheids. The linear circular rotation velocity
of the Sun around the center of the Galaxy is $V_0=239\pm7$ km
s$^{-1}$ for the accepted distance $R_0=8.0\pm0.2$~kpc, and the
Oort constants
 $A= 16.53\pm0.50$~km s$^{-1}$ kpc$^{-1}$ and
 $B=-13.34\pm0.68$~km s$^{-1}$ kpc$^{-1}$.
In comparison with the parameters specified in
the first column of the table~\ref{t:01}, the parameters~(\ref{Ceph1})
are determined with less errors, because the distance errors,
determined on the basis of the period-luminosity relation, are no more than
10--15\%.

By Bobylev~(2017) for this sample of Cepheids the following parameters were found
as a result of joint solution of three Bottlinger's equations (for $V_r, V_l,$ and $V_b$):
 $(U,V,W)_\odot=(7.9,11.7,7.4)\pm(0.7,0.8,0.6)$~km s$^{-1}$,
 $\Omega_0 =28.84\pm0.33$~km s$^{-1}$ kpc$^{-1}$,
 $\Omega{'}_0=-4.05\pm0.10$~km s$^{-1}$ kpc$^{-2}$ and
 $\Omega^{''}_0=0.81\pm0.07$~km s$^{-1}$ kpc$^{-3}$, where
$V_0=231\pm6$~km s$^{-1}$ ($R_0=8.0\pm0.2$~kpc),
 $A= 16.20\pm0.38$~km s$^{-1}$ kpc$^{-1}$ and
 $B=-12.64\pm0.51$~km s$^{-1}$ kpc$^{-1}$.
We see very good agreement with
solution~(\ref{Ceph1}), although errors
of the parameters found in the solution~(\ref{Ceph1}) are slightly larger due to
smaller number of equations used.

In paper by Mel'nik et al. (2015) were found the following parameters of Galactic rotation from 274 Cepheids with line-of-sight velocities taken from various sources and proper motions taken from
the HIPPARCOS catalog (ESA 1997; van Leeuwen 2007):
 $(U,V)_\odot=(8.1,12.7)\pm(0.8,1.0)$~km s$^{-1}$,
   $\Omega_0 =28.8\pm0.8$~km s$^{-1}$ kpc$^{-1}$,
 $\Omega{'}_0=-4.88\pm0.14$~km s$^{-1}$ kpc$^{-2},$
 $\Omega^{''}_0=1.07\pm0.17$~km s$^{-1}$ kpc$^{-3}$ and
$A=18.3\pm0.6$~km s$^{-1}$ kpc$^{-1}.$
Parameters~(\ref{Ceph1}) are in good agreement with the latter values. Number of used Cepheids
and equations are approximately the same in both cases, but thanks to the use of proper motions from the TGAS catalog, the kinematic parameters~(\ref{Ceph1}) are determined with smaller errors.

%%%%%%%%%%%%%%%%%%%%%%%%%%%%%%%%%%%%%%%%%%%%%%%%%%%%%%%%%%%%%% t2
 \begin{table}[t]
 \caption[]
  {\small
Parameters of Galactic rotation, found from OB stars with
the TGAS proper motions and trigonometric parallaxes
  }
  \begin{center}  \label{t:02}
  \small\protect
  \begin{tabular}{|l|r|r|r|r|}

   Parameter               &      $V_r,V_l$  &          $V_r$  &        $V_l$  \\\hline
   $U_\odot$               & $ 6.8\pm0.9$    & $ 6.4\pm1.7$    & $ 8.6\pm1.1$  \\
   $V_\odot$               & $ 9.0\pm1.0$    & $12.7\pm1.7$    & $ 4.7\pm1.2$  \\

 $\Omega_0$                & $29.82\pm0.62$  &            ---  & $30.58\pm0.56$  \\
 $\Omega{'}_0$            & $-4.34\pm0.14$  & $-4.68\pm0.25$  & $-4.29\pm0.15$  \\
 $\Omega_0^{''}$           & $ 0.67\pm0.07$  & $ 1.21\pm0.18$  & $ 0.50\pm0.07$  \\
   $\sigma_0$              &         $ 13.71$&       $16.05$   &        $10.97$  \\
               $N_\star$   &           $220$ &          $223$  &            $225$  \\
           $A$             & $ 17.35\pm0.56$ & $ 18.73\pm1.01$ & $ 17.18\pm0.60$ \\
           $B$             & $-12.48\pm0.83$ &             --- & $-13.40\pm0.82$ \\
 %\hline
 \end{tabular}\end{center}
% Note: $U_\odot, V_\odot, \sigma_0$ are in km s$^{-1}$, $\Omega_0, A, B$ are in km s$^{-1}$ kpc$^{-1}$,
% $\Omega{'}_0$ is km s$^{-1}$ kpc$^{-2}$,  $\Omega_0^{''}$ is in km s$^{-1}$ kpc$^{-3}$.
 \end{table}
%%%%%%%%%%%%%%%%%%%%%%%%%%%%%%%%%%%%%%%%%%%%%%%%%%%%%%%%%%%%%%%%% t2

\subsection{OB Stars}\label{OBstars}
In table~\ref{t:02} the parameters of Galactic rotation found from
OB stars with proper motions and parallaxes from the TGAS catalog
are given. We used the stars with relative errors of trigonometric
parallaxes $\sigma_\pi/\pi<200\%$. According to this table, we can
find the distance scale factor $p=(-4.68)/(-4.29)=1.09\pm0.07$,
those the TGAS distances should be increased on 9\%.
In this sample average distance ${\overline r}=1.9$~kpc and only
20 stars are located closer than 0.45~kpc from the Sun. Their
exclusion from the sample does not affect an estimate of the
coefficient $p.$

Using photometric distances for spectral-double stars and
``Calcium'' distances for OB stars, as well as proper motions from
the  TGAS catalog for these stars, the following kinematic
parameters were found as a result of joint solution of equations~(\ref{EQ1})--(\ref{EQ2}):
\begin{equation}
 \begin{array}{ccc}
 (U,V)_\odot=(7.5,9.5)\pm(0.8,1.1)~{\hbox {\rm km s$^{-1}$}},\\
     \Omega_0=30.82\pm0.59~{\hbox {\rm km s$^{-1}$ kpc$^{-1}$}},\\
 \Omega{'}_0=-4.50\pm0.12~{\hbox {\rm km s$^{-1}$ kpc$^{-2}$,}}\\
 \Omega_0^{''}=0.835\pm0.106~{\hbox {\rm km s$^{-1}$ kpc$^{-3}$.}}
 \label{Ceph-1}
 \end{array}
 \end{equation}
 Here, we used 244 OB stars, $V_0=247\pm8$~km s$^{-1}$
($R_0=8.0\pm0.2$~kpc), the Oort constants
 $A= 17.98\pm0.51$~km s$^{-1}$ kpc$^{-1}$ and
 $B=-12.83\pm0.78$~km s$^{-1}$ kpc$^{-1}$.

In Bobylev \&~Bajkova~(2017) the following kinematic parameters were found when using a smaller
number of stars from this sample and solving three equations (for $V_r, V_l,$ and $V_b$):
 $(U,V,W)_\odot=(8.2,9.3,8.8)\pm(0.7,0.9,0.7)$~km s$^{-1}$,
 $\Omega_0=31.53\pm0.54$~km s$^{-1}$ kpc$^{-1}$,
 $\Omega{'}_0=-4.44\pm0.12$~km s$^{-1}$ kpc$^{-2}$,
 $\Omega^{''}_0=0.706\pm0.100$~km s$^{-1}$ kpc$^{-3}$,
wherein $V_0=252\pm8$~km s$^{-1}$
($R_0=8.0\pm0.2$~kpc), $A=-17.77\pm0.46$~km s$^{-1}$ kpc$^{-1}$ and
$B=13.76\pm0.71$~km s$^{-1}$ kpc$^{-1}$.

Note that Bobylev~\&~Bajkova (2017) found the value $p=1.04$ from kinematic analysis of OB stars with
proper motions and  parallaxes from the TGAS catalog.

 \section{CONCLUSIONS}
Galactic rotation parameters are determined using three
samples of young stars with different distance scales. These samples
were studied previously using data from various
catalogs. The first sample contains massive spectral-double
OB stars with photometric distances, the second one consists of OB stars,
which distances are determined along the lines of interstellar Calcium,
the third sample consists of classical Cepheids, the distances to which are determined
using the period-luminosity relation. In this paper, we use those stars from these samples that are included in the Gaia TGAS catalog.

In this paper, in contrast to the papers of Bobylev~(2017) and Bobylev~\&~Bajkova~(2017), only two main
kinematic equations with left-hand sides $V_r$ and $V_l$ are considered. From 250
Cepheids, the distances to which are calculated on the basis of the period--luminosity relation, and the proper motions taken from the TGAS catalog, the following kinematic parameters are found
as a result of joint solution of the equations for $V_r$ and $V_l$:
 $(U,V)_\odot=(7.9,11.0)\pm(0.8,1.0)$~km s$^{-1}$,
     $\Omega_0=29.87\pm0.45$~km s$^{-1}$ kpc$^{-1}$,
 $\Omega{'}_0=-4.13\pm0.13$~km s$^{-1}$ kpc$^{-2}$,
 $\Omega_0^{''}=0.50\pm0.12$~km s$^{-1}$ kpc$^{-3}$.
Here the circular rotation velocity of the Sun around the center
of the is $V_0=239\pm7$ km s$^{-1}$ (for the accepted distance
$R_0=8.0\pm0.2$~kpc), and the Oort constants
 $A= 16.53\pm0.50$~km s$^{-1}$ kpc$^{-1}$ and
 $B=-13.34\pm0.68$~km s$^{-1}$ kpc$^{-1}$.

Based on a similar approach for 244 OB stars, the following
kinematic parameters are found:
 $(U,V)_\odot=(7.5,9.5)\pm(0.8,1.1)$~km s$^{-1}$,
     $\Omega_0=30.82\pm0.59$~km s$^{-1}$ kpc$^{-1}$,
 $\Omega{'}_0=-4.50\pm0.12$~km s$^{-1}$ kpc$^{-2}$,
 $\Omega_0^{''}=0.835\pm0.106$~km s$^{-1}$ kpc$^{-3}$.
 Here $V_0=247\pm8$~km s$^{-1}$,
 $A= 17.98\pm0.51$~km s$^{-1}$ kpc$^{-1}$ and
 $B=-12.83\pm0.78$~km s$^{-1}$ kpc$^{-1}$.

According to the samples of Cepheids and OB stars, kinematic
parameters were determined using also trigonometric parallaxes. In this case we used the stars with
relative errors of parallaxes less than 200\%. From the comparison
of values $\Omega{'}_0$, found as a result of solutions of equations for $V_r$ and $V_l$ separately,
the distance scale coefficient $p$ has been determined. Here, we
start from the assumption that the kinematic parameters, found from line-of-sight
velocities, are less  dependent on distance errors than those found
from proper motions. According to Cepheids, we found that the distance scale of the TGAS
catalog should be reduced by
3\% ($p=0.97\pm0.07$), and for OB stars, on the contrary, it should be
increased by 9\% ($p=1.09\pm0.07$).

\section{Acknowledgment}
We are grateful to the referee for the helpful remarks that
contributed to an improvement of this paper. This study was
supported by the ``Transient and explosive processes in
astrophysics'' Program of the Presidium of Russian Academy of
Sciences (P--7).

\section{References}

\begin{itemize}

\item Benedict G.F., McArthur B.E., Nelan E.P., and Harrison T.E. 2017, PASP, 129, 2001

\item Bobylev V.V., and Bajkova A.T. 2011, Astron. Lett., 37, 526   %CaII--2.

\item Bobylev V.V., and Bajkova A.T. 2014, MNRAS, 437, 1549

\item Bobylev V.V., and Bajkova A.T. 2015, Astron. Lett., 41, 473   %OB-SB--2; CaII--2.

\item Bobylev V.V., 2017, Astron. Lett., 43, 152

\item Bobylev V.V., and Bajkova A.T. 2017, Astron. Lett., 43, 159

\item Gaia Collaboration, Brown A.G.A., Vallenari A., Prusti T., et al. 2016, A\&A, 595, 2

\item Casertano S., Riess A.G., Bucciarelli B., and Lattanzi M.G. 2017, A\&A, 599, 67

\item Clementini G., L. Eyer L., V. Ripepi V., et al. 2017, A\&A, 605, 79

\item The HIPPARCOS and Tycho Catalogues, 1997, ESA SP--1200

\item Jao W.-C., Henry T.J., Riedel A.R., et al. 2017, ApJ, 832, L18

\item Lindegren L., Lammers U., Bastian U., et al. 2016, A\&A, 595, 4

\item Mel'nik A.M., Rautiainen P., Berdnikov L.N., et al. 2015, AN, 336, 70

\item Gaia Collaboration, Prusti T., de Bruijne J.H.J., Brown A.G.A., et al. 2016, A\&A, 595, 1

\item Rastorguev A.S., Zabolotskikh M.V., Dambis A.K., et al. 2017, Astrophys. Bulletin, 72, 122

\item Sch\"onrich R., and Aumer M., 2017, MNRAS, 472, 3979

\item Stassun K.G., and Torres G. 2016, ApJ, 831, 74

\item Vall\'ee J.P. 2017, Astrophys. Space Sci., 362, 79

\item van Leeuwen F. 2007, A\&A, 474, 653

\item Zabolotskikh M.V., Rastorguev A.S., Dambis A.K., et al. 2002, Astron. Lett. 28, 454

\end{itemize}

\end{document}